\documentclass[aps,pra,twocolumn,superscriptaddress]{revtex4-1}

\usepackage{graphicx}
\usepackage{multirow}
\usepackage{epstopdf}
\usepackage{array}
\usepackage{longtable}
\usepackage{rotating,booktabs}
\usepackage{booktabs,threeparttable}
\usepackage{bm}
\usepackage{CJK}
\usepackage[marginal]{footmisc}
\usepackage{amsmath}
\usepackage{float}
\makeatletter

\newcommand{\Rmnum}[1]{\expandafter\@slowromancap\romannumeral #1@}
\makeatother
\usepackage{setspace}

\usepackage{color}

\begin{document}


\title{Collisional-radiative modeling of the $5p - 5s$ spectrum of W XIV - W XVI ions}


\author{Xiaobin Ding}
\email{dingxb@nwnu.edu.cn}
\affiliation{Key Laboratory of Atomic and Molecular Physics and Functional Materials of Gansu Province, College of Physics and Electronic Engineering, Northwest Normal University, Lanzhou 730070, China}

\author{Fengling Zhang}
\affiliation{Key Laboratory of Atomic and Molecular Physics and Functional Materials of Gansu Province, College of Physics and Electronic Engineering, Northwest Normal University, Lanzhou 730070, China}
\author{Yang Yang}
\affiliation{Key Laboratory of Applied Ion Beam Ministry of Education, Institute of Modern Physics, Fudan University, Shanghai 200433, China}
\author{Ling Zhang}
\affiliation{Institute of Plasma Physics, Chinese Academy of Sciences, Hefei 230000, China}
\author{Fumihiro Koike}
\affiliation{Department of physics, Sophia University, Tokyo 102-8554, Japan}
\author{Izumi Murakami}
\affiliation{National Institute for Fusion Science, National Institutes of Natural	Sciences, Toki, Gifu 509-5292, Japan}
\affiliation{Department of Fusion Science, The Graduate University for Advanced Studies, SOKENDAI, Toki, Gifu 509-5292, Japan}
\author{Daiji Kato}
\affiliation{National Institute for Fusion Science, National Institutes of Natural	Sciences, Toki, Gifu 509-5292, Japan}
\affiliation{Department of Advanced Energy Enginnering Science, Kyushu University, Fukuoka, 816-8580, Japan}
\author{Hiroyuki A Sakaue}
\affiliation{National Institute for Fusion Science, National Institutes of Natural	Sciences, Toki, Gifu 509-5292, Japan}
\author{Nobuyuki Nakamura}
\affiliation{Institute for Laser Science, The University of Electro-Communications, Chofu, Tokyo 182-8585, Japan}
\author{Chenzhong Dong}
\affiliation{Key Laboratory of Atomic and Molecular Physics and Functional Materials of Gansu Province, College of Physics and Electronic Engineering, Northwest Normal University, Lanzhou 730070, China}


\date{\today}

\begin{abstract}
The wavelength and rate of the $5p-5s$ transition of W XIV - W XVI ions have been calculated by the relativistic configuration interaction (RCI) method with the implementation of Flexible Atomic code (FAC). A reasonable collisional-radiative model (CRM) has been constructed to simulate the $5p - 5s$ transition spectrum of W XIV - W XVI ions which had been observed in electron beam ion trap (EBIT) device. The results are in reasonable agreement with the available experimental and theoretical data, and might be applied to identify the controversial spectra. The confusion on the assignment of the ionization stage are solved in the present work. 
\end{abstract}

\pacs{32.30.−r,31.15.am,32.70.Fw}
\maketitle

\maketitle
  

\section{INTRODUCTION} 

Tungsten is selected as the divertor and plasma-facing material for magnetic confinement fusion devices, such as ITER and EAST, due to its high melting point and high sputtering energy threshold, low sputtering rate and low deuterium/tritium retention rate\cite{1,2}. However, due to the interaction between the edge plasma and the wall material, tungsten might be ionized and transported to the high temperature core plasma region and be ionized further. Therefore, tungsten ion may exist as the intrinsic impurities both in the plasma core and edge region. Large radiation power loss caused by highly charged tungsten ions will lead to the degradation of the plasma performance or even plasma disruption if the relative concentration of tungsten ions in the core plasma is higher than $10^{-5}$\cite{56}. Quantitative diagnostics of tungsten influx and concentration are very important to understand tungsten transport in fusion plasma. Furthermore the control of tungsten impurity is one of the key issues of the steady-state operation of future fusion reactor. On the other hand, a tiny amount of tungsten ion is beneficial to diagnose the plasma parameters such as temperature and density of fusion plasma.  Therefore, the knowledge of the atomic structure, properties, kinetic process and emission spectra of various tungsten ions are required and are expected to have great significance for the diagnosis of fusion plasma. 

The highly ionized tungsten ion have relative simple structure, and had been widely studied\cite{5,6,8,9,10,11,14,15,16,26}. For the tungsten ions with lower degree of ionization, many studies on its spectral line can also be found\cite{17,18,19,20,21,22,23,24,29,30,32}. However, the data of tungsten ion with moderate ionization degree is still relatively lacking, especially for the data of W VIII - W XXVIII ions\cite{3,31}. These tungsten ions involves several open 4$f$ shell, which makes the spectrum not well isolated and become complex to explain. Meanwhile, the theoretical calculation on the atomic structure and spectrum are complicated due to the influence of the 4$f$ wave function collapse. In addition, relativistic effects and electron-correlation effects in these ions have important influence on their structure and transition properties, especially on their ground and low-lying excited state\cite{35,39,40}.

Electron beam ion trap (EBIT) is now being widely used to observe the spectrum of the tungsten ions\cite{22,29,46,47,49,58,59,63}. By observing the dependence of the spectrum intensity on the incident electron beam energy, the spectrum could be assigned to the appropriate ions. Therefore, it provides a plenty of knowledge on the transitions and atomic properties of tungsten ions.

In 2015, W. Li et al. measured the EUV spectra of W$^{11+}$ - W$^{15+}$ on Shanghai EBIT\cite{42}. They assigned the lines in 24.62 - 25.22 nm, 24.32 - 24.92 nm, 23.27 - 24.09 nm, 22.54 - 23.37 nm, and 21.48 - 22.69 nm to $5p - 5s$ transition of W$^{11+}$ - W$^{15+}$ ions, respectively. In the same year, Y. Kobayashi et al. measured the EUV and visible spectra of W$^{12+}$ - W$^{14+}$ ions on Tokyo EBIT\cite{43}. They thought that the lines at 24.32 nm, 24.77 nm, 24.83 nm and 24.91 nm are the $5p - 5s$ transition of W$^{13+}$ ion, while the lines in 23.27 - 24.09 nm are from W$^{14+}$ ion. The identification on the ionization degree, from these two independent works are different by one ionization state. To clarify the confusion on the identification, the spectrum of the $5p - 5s$ of W$^{13+}$ - W$^{15+}$ ions was calculated by collisional-radiative model in this paper. 
 

\section{Theoretical method}

Collisional-radiative model (CRM) has been widely used to simulate and explain the observed plasma spectrum\cite{4,41,42,38,46,37,47,36,49,51,22,29,54,57}. The spectral intensity $I_{p,q}(\lambda)$ of a transition with wavelength $\lambda$ from the upper excited level $p$ to the lower level $q$ can be expressed by the equation: 
\begin{eqnarray}\label{eq1}
   I_{p,q}(\lambda) \propto n(p) A(p,q)\phi(\lambda).  
\end{eqnarray}
where, $A(p,q)$ is the radiative transition rate or Einstein coefficient of the transition from $p$ to $q$, which can be obtained accurately by experimental observation or theoretical calculation. The function $\phi(\lambda)$ is the normalized line profile, which was taken as Gaussian profile to include the Doppler, natural, collisional and instrumental broadening effects in the present work. $n(p)$ is the population of the upper excited level $p$, which was determined by the atomic processes in the plasma and can be obtained by solving the rate equation. To construct the rate equation, the most important atomic processes in the plasma, such as spontaneous radiation transitions, collision and de-excitation, impact ionization, radiation recombination, and three-body recombination etc. should be taken into account.

In the EBIT, the electron beam energy is assumed to be mono-energetic distribution. The ionization degree of the ions generated in the ion trap is relatively simple\cite{44}. The low-density plasma in the EBIT can be regarded as optically thin and isotropic, and its ionization and recombination process are much slower than the collisional and radiative processes. To simulate the specific ions, the impact ionization, radiation recombination, three-body recombination and charge exchange processes are ignored in the present work. The following rate equation can be used to describe the population of the excited upper levels $p$:
 \begin{eqnarray}\label{eq2}
   \begin{split}
        \frac {d}{dt}n(p)&=\sum_{q>p}F(q,p)n_en(q)+\sum_{p<q}[C(q,p)n_e+A(q,p)]n(q) \\
                               &-[\sum_{q>p}C(p,q)n_e+\sum_{q<p}F(p,q)n_e+\sum_{p>q}A(p,q)]n(p) \\                            
   \end{split}                         
 \end{eqnarray}
where, ${n_e}$ is the electron density of the plasma, $C(p,q)$ and $F(q,p)$ are collisional excitation and de-excitation rates coefficient from the level $p$ to $q$, respectively. These rate coefficients can be obtained by convoluting the cross section of the collision (de)excitation with the free electron energy distribution function, which can be described by the $\delta$ function for mono-energetic electron beam of EBIT. The collision excitation cross section can be obtained by the distorted wave approximation, and the collision de-excitation cross section can be obtained according to the principle of the detailed balance. The first and second terms in the righthand side of Eq.~(\ref{eq2}) refer to the population flux from the other energy levels to the  level $p$, and the third term represents the depopulation flux from the level $p$ to the other levels. The rate equation can be solved in the Quasi-Steady State (QSS) approximation $\frac {d}{dt}n(p)=0$. 

For heavy ions such as tungsten (Z=74), relativistic effects and electron-correlation effects have important influence on its structure and transition properties. Therefore, the relativistic configuration interaction method (RCI) was used with the implementation of flexible atomic code(FAC)\cite{52}. The atomic data including the energy levels, radiative transition rates, and cross sections of collisional excitation are calculated. The configurations included in the CRM and the correlation configurations of W$^{13+}$ - W$^{15+}$ ions are given in Table~\ref{Tab1}.

\begin{table}
\centering
\caption{Configurations included in the CRM and Correlation configurations of W$^{13+}$ - W$^{15+}$.}\label{Tab1}
\begin{tabular}{lll}
\hline\hline
 W$^{13+}$ & W$^{14+}$  &  W$^{15+}$       \\
\hline
\multicolumn{3}{c}{ Configurations included in the CRM}  \\
$4f^{13}5s^2$ & $4f^{12}5s^2$  & $4f^{11}5s^2$ \\
$4f^{13}5s5p$ & $4f^{12}5s5p$  & $4f^{11}5s5p$ \\
$4f^{13}5s5d$ & $4f^{11}5s^25p$   & $4f^{10}5s^25p$ \\
$4f^{12}5s^25p$ \  \  \ \  \ \ \ \ \ \ \   &  $4f^{11}5s5p^2$  \  \ \ \ \ \  \  \  \ \ \  &    $4f^{11}5s^25d$ \ \   \  \  \\
$4f^{12}5s^25d$ &      &    \\
$4f^{12}5s5p^2$ &      &    \\
\hline
\multicolumn{3}{c}{ Correlation configurations}  \\
$4f^{14}5s$ & $4f^{12}5s5d$  & $4f^{12}5s$ \\
$4f^{14}5p$ & $4f^{11}5s^25d$  & $4f^{11}5p^2$ \\
$4f^{14}5d$ & $4f^{13}5s$  & $4f^{11}5p5d$ \\
$4f^{13}5s5f$ & $4f^{13}5p$  & $4f^{11}5s5d$ \\
$4f^{13}5s5g$ & $4f^{13}5d$  & $4f^{11}5s6p$ \\
$4f^{13}5s5d$ & $4f^{12}5p5d$   & $4f^{11}5s6f$ \\
$4f^{13}5p^2$ & $4f^{12}5f^2$ & $4f^{10}5s^26d$ \\
$4f^{13}5p5d$ & $4f^{10}5s^25f^2$   & $4f^{12}5p$   \\
$4f^{13}5p5f$ & $4f^{14}$  & $4f^{12}5d$   \\
$4f^{13}5p5g$ &    &   \\
$4f^{13}5d^2$ &    &   \\
$4f^{13}5d5f$ &     &   \\
$4f^{13}5d5g$ &   &   \\
$4f^{13}5f^2$ &      &    \\
$4f^{12}5f5g$ &      &    \\
$4f^{13}5g^2$ &      &    \\
\hline\hline
\end{tabular}
\end{table}

\section{Results and discussion}

\subsection{The transition and spectrum of W$^{13+}$ ion}

The ground configuration of W$^{13+}$ ion is $4f^{13}5s^2$, which splits into $[(4f^5_{5/2})_{5/2}5s^2]_{5/2}$ and $[(4f^7_{7/2})_{7/2}5s^2]_{7/2}$ doublet levels, while $[(4f^7_{7/2})_{7/2}5s^2]_{7/2}$ is the ground state. The notations given in here are in relativistic form with the full relativistic orbital omitted. The excitation energy of the first excited state is presented in Table~\ref{Tab2} compared with the existing data from the calculation and experiment. The results calculated using the Multi-reference Model potential (MR-MP) theory by  M. J. Vilkas et al. is quite different from both the experiment observation and the theoretical calculation, mainly because the former is the theory using empirical model potentials. Such deviations might be due to inappropriate choices of the experimental values for the model potential like we found in the previous work\cite{55}. The present RCI calculation is in good agreement with the results both experimental and calculated by Z. Zhao et al.'s\cite{41}. There is little difference between Y. Kobayashi et al.'s result and others. It might caused by inappropriate assignment of the M1 transitions in their work. Calculation on the spectrum of M1 transition of W$^{13+}$ ion is also in progress, and will be published elsewhere.

\begin{table}
	\footnotesize
	\centering
	\caption{Excitation energy (in eV) of the first excited state $[(4f^5_{5/2})_{5/2}5s^2]_{5/2}$ in W$^{13+}$.}\label{Tab2}
	\begin{tabular}{lllll}
		\hline\hline
		Level  \  \  \  \  \  \   \  \  & Present  \  \  \  \ & MR-MP\cite{40}  \  \  \  \  & Exp\cite{41} \  \  \  \ & Exp\cite{43}  \  \  \\
		\hline
		$(4f^5_{5/2})_{5/2}$ & 2.2467 & 3.1896  & 2.2545 & 2.2130  \\
		\hline\hline
	\end{tabular}
\end{table}

The calculated transition wavelength and transition rates of $5d - 5p$ and $5p - 5s$ transitions of W$^{13+}$ ion are presented in Table~\ref{Tab3} with the values of other theories and experiment. For the transition $5d - 5p$, only the transition with large transition rate are provided.

\begin{table*}
\footnotesize
\centering
\caption{Wavelength $\lambda$ (in nm) and the transition rate $A_{(p,q)}$ (in $10^{11}s^{-1}$) of $5d - 5p$ and $5p - 5s$ of W$^{13+}$ ion. The column `Key' correspond to the label in Fig.~\ref{fig1}.}\label{Tab3}
\begin{threeparttable}
\begin{tabular}{cllllllll}
\midrule
\hline
\hline
 \  \ Key \  \ & Lower    &  Upper    & $\lambda$ &     &   & $\lambda_{exp}$\cite{43} \  \   & $A_{(p,q)}$ &      \\
\hline	
   & $[(4f^5_{5/2}4f^7_{7/2})_65p_{1/2}]_{13/2}$  \  \  \  \ \  \   & $[(4f^5_{5/2}4f^7_{7/2})_65d_{3/2}]_{15/2}$          & 18.18  \  \  \  \   &                   &                  &           & 1.73  &                  \\
   & $[(4f^6_{7/2})_65p_{3/2}]_{15/2}$                  & $[(4f^6_{7/2})_65d_{5/2}]_{17/2}$                         & 21.03 &                   &                  &           & 1.52  &                 \\
   & $[(4f^5_{5/2}4f^7_{7/2})_55p_{3/2}]_{13/2}$  & $[(4f^5_{5/2}4f^7_{7/2})_55d_{5/2}]_{15/2}$         & 21.05  &                   &                  &           & 1.53  &                 \\
   & $[(4f^4_{5/2})_{4}5p_{3/2}]_{11/2}$                & $[(4f^4_{7/2})_45d_{5/2}]_{13/2}$                         & 21.07 &                   &                  &           & 1.53  &                  \\
   & $[(4f^5_{5/2}4f^7_{7/2})_55p_{3/2}]_{15/2}$  & $[(4f^5_{5/2}4f^7_{7/2})_65d_{5/2}]_{17/2}$         & 21.13  &                   &                  &           & 1.50  &                  \\
1 & $[(4f^5_{5/2})_{5/2}5s^2]_{5/2}$                    & $[((4f^5_{5/2})_{5/2}5s_{1/2})_35p_{3/2}]_{5/2}$  \  \  \  \  \  \   & 24.09  & $23.87^c \  \  \  \   $  & $24.00^d$  \  \  \  \  &           & 0.62  & $0.63^c$  \  \  \  \  \\
2 & $[(4f^7_{7/2})_{7/2}5s^2]_{7/2}$                    & $[((4f^7_{7/2})_{7/2}5s_{1/2})_45p_{3/2}]_{7/2}$  & 24.17  & $23.95^c$  & $24.06^d$ & 24.32 & 0.55  & $0.54^c$  \\   
3 & $[(4f^5_{5/2})_{5/2}5s^2]_{5/2}$                    & $[((4f^5_{5/2})_{5/2}5s_{1/2})_25p_{3/2}]_{7/2}$  & 24.61  & $24.41^c$  & $24.57^d$ & 24.77 & 0.53  & $0.51^c$  \\
4 & $[(4f^7_{7/2})_{7/2}5s^2]_{7/2}$                    & $[((4f^7_{7/2})_{7/2}5s_{1/2})_35p_{3/2}]_{9/2}$  & 24.69  & $24.53^c$  & $24.64^d$ & 24.83 & 0.55  & $0.61^c$  \\
5 & $[(4f^7_{7/2})_{7/2}5s^2]_{7/2}$                    & $[((4f^7_{7/2})_{7/2}5s_{1/2})_45p_{3/2}]_{5/2}$  & 24.76  & $24.71^c$  & $24.70^d$ & 24.91 & 0.57  & $0.54^c$  \\
6 & $[(4f^5_{5/2})_{5/2}5s^2]_{5/2}$                    & $[((4f^5_{5/2})_{5/2}5s_{1/2})_35p_{3/2}]_{3/2}$  & 24.93  &                   &                  &            &0.58  &                  \\
\hline
\hline
\end{tabular}
\begin{tablenotes}
\item[] $^c$ From Y. Kobayashi et al with HULLAC code\cite{43}.
\item[] $^d$ From U. Safronova et al with Hartree-Fock-relativistic method (COWAN code)\cite{53}.
\end{tablenotes}
\end{threeparttable}
\end{table*}

The calculated transition with wavelengths of 17.0 - 22.00 nm dominated by $5d - 5p$ from $4f^{12}5s^25d$ to $4f^{12}5s^25p$. These transition have large transition rate, but have not been observed in EBIT experiment\cite{43}. The observed spectrum by EBIT in the wavelengths of 24.00 - 25.00 nm corresponds to the calculated $5p - 5s$ transitions from $4f^{13}5s5p$ to $4f^{13}5s^2$. Compared the calculated wavelength with the four experimental lines measured by Y. Kobayashi et al. in EBIT, the present calculation discrepancies are 0.62\%, 0.65\%, 0.56\% and 0.60\%; Y. Kobayashi et al.\cite{43} are 1.52\%, 1.45\%, 1.21\% and 0.80\% and U.Safronova et al.\cite{53} are 1.07\%, 0.81\% and 0.77\%, 0.84\%, respectively. By comparison, it can be found that the current calculation results make better agreement with the experimental values.

\begin{figure*}
\centering
\includegraphics{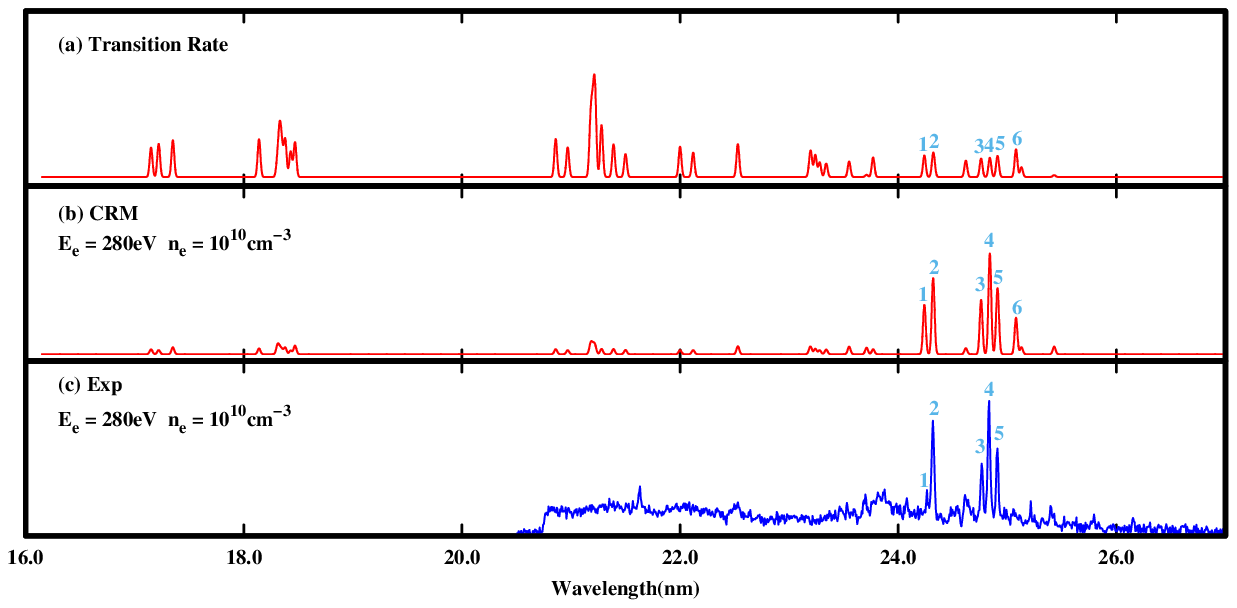}
\caption{Comparison between the experimental and calculated spectra of W$^{13+}$ ion. (a) The calculated radiative transition rate (Shifted to the right by 0.15 nm); (b) The spectral intensity calculated by the CRM (Shifted to the right by 0.15 nm); (c) Experimental spectra measured by Y. Kobayashi et al on the EBIT device with the electron density $n_e = 10^{10}cm^{-3}$ and the energy of electron beam $E_e = 280eV$.}\label{fig1}
\end{figure*}

The calculated and experimental spectra of W$^{13+}$ ion in the range of 17.0 - 27.0 nm is shown in Fig.~\ref{fig1}. The upper panel Fig.~\ref{fig1} (a) is the calculated radiative transition rate; the middle panel Fig.~\ref{fig1} (b) is the spectral intensity calculated by the CRM with the electron density $n_e = 10^{10}cm^{-3}$ and the incident electron beam energy $E_e = 280eV$; and the bottom panel Fig.~\ref{fig1} (c) is the experimental spectra measured  by Y. Kobayashi et al on the EBIT device with the electron density $n_e = 10^{10}cm^{-3}$ and the electron beam energy $E_e = 280eV$\cite{43}. In the figure, each individual transition was assumed to have the Gaussian profile with full width at half maximum (FWHM) 0.03 nm, which correspond to the experimental resolution. The theoretical transition wavelengths are shifted to longer wavelength by 0.15 nm to compare with the experimental spectra. The peaks in wavelengths of 23.00 - 23.62 nm are mainly from $4f^{12}5s5p^2$ to $4f^{12}5s^25p$ transition and the wavelengths of 24.00 - 25.00 nm are from $4f^{13}5s5p$ to $4f^{13}5s^2$ transition. As can be seen from figure, the transition near 20.86 - 21.17 nm have large rate (Fig.~\ref{fig1} (a)), but it has not been observed in the experiment in Fig.~\ref{fig1} (c). This might caused by the population mechanism of the excited upper levels.  

In order to analyze the spectral intensity, the transition rates, population flux and intensity for some selected transitions are given in Table~\ref{Tab4}. For convenience, the population flux from collisional de-excitation of the higher levels was given by $I_{Fin}=\sum_{q>p}F(q,p)n_en(q)$, the population flux from collisional excitation of the lower levels was given by $I_{Cin}=\sum_{p<q}C(q,p)n_en(q)$ and population flux from spontaneous radiation transition of the higher levels was given by $I_{Ain}=\sum_{p<q}A(q,p)n(q)$, respectively.

\begin{table*}
\centering
\caption{The transition rate $A_{(p,q)}$ (in $10^{11}s^{-1}$),  population flux ($\sum_{q>p}F(q,p)n_en(q)$, $\sum_{p<q}C(q,p)n_en(q)$ and $\sum_{p<q}A(q,p)n(q)$ represented by $I_{F_{in}}$, $I_{C_{in}}$ and $I_{A_{in}}$, respectively) and intensity Int (in $10^{-2}cm^{-3}s^{-1}$) of $5d - 5p$ and $5p - 5s$ of W$^{13+}$ ion. $I_{C_{in}}$ and $I_{A_{in}}$ are  in $10^{-2}cm^{-3}s^{-1}$, and $I_{F_{in}}$ in $10^{-13}cm^{-3}s^{-1}$.}\label{Tab4}
\begin{tabular}{llllllll}
\hline\hline
 \ \ \ $\lambda$ & Upper &  $A_{(p,q)}$  & $I_{C_{in}}$ & $I_{A_{in}}$  & $I_{F_{in}}$  \  \     \  \   &  Int  \  \     \\
\hline
 \ \ \ 18.18  \  \  \  \  \  \  \  \  & $[(4f^5_{5/2}4f^7_{7/2})_65d_{3/2}]_{15/2}$         & 1.26 & 6.53 & 7.73$\times 10^{-12}$ \  \   & 1.47 & 4.77 \\
 \ \ \ 21.03 & $[(4f^6_{7/2})_65d_{5/2}]_{17/2}$                         & 1.52 & 9.48 & 0             & 3.21 & 9.48 \\
 \ \ \ 21.05 & $[(4f^5_{5/2}4f^7_{7/2})_55d_{5/2}]_{15/2}$         & 1.49 & 6.87 & 0             & 5.74 & 6.66 \\
 \ \ \ 21.07 & $[(4f^4_{7/2})_45d_{5/2}]_{13/2}$                         & 1.41 & 4.27 & 0             & 2.53 & 3.93 \\
 \ \ \ 21.13 & $[(4f^5_{5/2}4f^7_{7/2})_65d_{5/2}]_{17/2}$         & 1.50 & 4.90 & 0             & 1.03 & 4.90 \\
 \ \ \ 24.09 & $[((4f^5_{5/2})_{5/2}5s_{1/2})_35p_{3/2}]_{5/2}$  \  \ \  \  \   \  \  \  \  \  \  & 0.62  \  \   \  \    \  \  \  \  \  \ & 39.8  \  \  \  \   \  \  \  \  \  \ & 8.26  \  \  \  \   \  \  \  \  \  \   & 39.3 \  \   \  \  \   \  \  \  \  & 46.7  \  \  \  \   \\
 \ \  \ 24.17 & $[((4f^7_{7/2})_{7/2}5s_{1/2})_45p_{3/2}]_{7/2}$  & 0.55 & 73.8 & 3.64        & 28.9 & 69.4 \\
 \ \ \ 24.61 & $[((4f^5_{5/2})_{5/2}5s_{1/2})_25p_{3/2}]_{7/2}$  & 0.53 & 55.5 & 2.50        & 36.2 & 51.5 \\
 \ \ \ 24.69 & $[((4f^7_{7/2})_{7/2}5s_{1/2})_35p_{3/2}]_{9/2}$  & 0.55 & 91.6 & 4.52        & 61.3 & 95.4 \\
 \ \ \ 24.76 & $[((4f^7_{7/2})_{7/2}5s_{1/2})_45p_{3/2}]_{5/2}$  & 0.57 & 58.0 & 2.07        & 33.2 & 59.0 \\
 \ \ \ 24.93 & $[((4f^5_{5/2})_{5/2}5s_{1/2})_35p_{3/2}]_{3/2}$  & 0.68 & 27.4 & 3.44        & 24.9 & 30.6 \\
\hline\hline
\end{tabular}
\end{table*}
 
As can be seen from Table~\ref{Tab4}, the collisional de-excitation flux $I_{F_{in}}$ are generally smaller than the other two processes by 11 orders of magnitude so that can be ignored. The population is mainly from the collisional excitation and the spontaneous radiative transition. For the excited $5d$ upper level, its population mainly comes from the collisional excitation of the lower level. And the spontaneous radiative transition from the upper level is small, because the population of the higher  level is very small or even zero. For the excited $5p$ levels, its population mainly comes from the collisional excitation of the lower level and the spontaneous radiation transition of the upper level and $I_{A_{in}}$ can not be ignored. The collisional excitation flux $I_{C_{in}}$ of the upper level of $5d$ is smaller than that of $5p$. As a result, the population of $5d$ upper level is smaller than the population of $5p$ upper level. This is the reason why those lines with large transition rate could not be observed in the EBIT experiment. 

We also calculated the spectra at the same electron beam energy as in the experiment of W. Li et al.\cite{42} and found that the relative intensity of the spectrum of W$^{13+}$ is insensitive to the electron beam energy. It is found that the present calculated W$^{13+}$ spectrum agrees well with the Y. Kobayashi et al.\cite{43} EBIT observation, as well as the observation by W. Li et al.\cite{42}, except the assignment of W. Li et al.'s work is W$^{12+}$ instead of W$^{13+}$ ion.


\subsection{The transition and spectrum of W$^{14+}$ ion}

The calculated ground configuration of W$^{14+}$ ion is $4f^{12}5s^2$. The transition wavelength, transition rate and intensities of $5p - 5s$ of W$^{14+}$ ion are shown in Table~\ref{Tab5}. The transition in wavelength of 22.50 - 24.50 nm are $5p - 5s$ from $4f^{12}5s5p$ to $4f^{12}5s^2$ and $4f^{11}5s5p^2$ to $4f^{11}5s^25p$. In these two transitions, the transition rate of the former is larger than that of the latter by about two orders of magnitude. The experimental observations of W. Li and Y. Kobayashi both have 9 peaks with high intensity in the wavelength range of 22.50 - 24.50 nm. In fact, there are more than 30 transitions with high transition rate obtained by the present calculation. In the table, only the transition data of the most intense are given. These lines are close in wavelength and they have similar intensities. For example, the peaks with wavelength 23.40 nm and 23.69 nm are both blended. In fact, the observed peaks with the key 4, 6 and 11 have two components mixed together, and the lines with the key 10 and 13 have three components mixed together\cite{42,43}.

\begin{table*}
\centering
\caption{Wavelength $\lambda$ (in nm), transition rate $A_{(p,q)}$ (in $10^{10}s^{-1}$) and intensity Int (in $cm^{-3}s^{-1}$) from $4f^{12}5s5p$ to $4f^{12}5s^2$ transitions in W$^{14+}$ ion. The column `Key' correspond to the label in Fig.~\ref{fig2}.}\label{Tab5} 
\begin{tabular}{clllll}
\hline\hline
 \  \   Key \  \   & Lower & Upper & $\lambda$ & $A_{(p,q)}$ & Int \\
\hline
 \  \   1  \  \     & $[(4f^5_{5/2}4f^7_{7/2})_{5/2}5s^2]_6$  \  \   \  \  \  \  \  \  \  \  & $[((4f^5_{5/2}4f^7_{7/2})_65s_{1/2})_{13/2}5p_{3/2}]_6$  \  \   \  \   \  \  \  \  \  \    &22.83 \  \  \  \  \  \  \  \  \  \  &8.29 \  \   \  \  \  \ \  \  \  \  \ &2.78 \  \  \  \   \\
 \  \   2   \  \    &$[(4f^5_{5/2}4f^7_{7/2})_{5/2}5s^2]_5$          & $[((4f^5_{5/2}4f^7_{7/2})_55s_{1/2})_{11/2}5p_{3/2}]_5$           &23.19&3.32&1.20 \\
 \  \   3  \  \     & $[(4f^5_{5/2}4f^7_{7/2})_{5/2}5s^2]_5$         & $[((4f^5_{5/2}4f^7_{7/2})_45s_{1/2})_{7/2}5p_{3/2}]_5$             &23.28&4.01&1.47 \\
 \  \   4  \  \     & $[(4f^6_{7/2})_65s^2]_6$                              & $[((4f^6_{7/2})_65s_{1/2})_{11/2}5p_{3/2}]_6$                           &23.32&6.12&3.62 \\
        &$[(4f^4_{5/2})_25s^2]_2$                               & $[((4f^4_{5/2})_25s_{1/2})_{3/2}5p_{3/2}]_3$                            &23.34&6.84&1.07 \\
 \  \   5  \  \     &$[(4f^6_{7/2})_25s^2]_2$                               & $[((4f^6_{7/2})_25s_{1/2})_{3/2}5p_{3/2}]_3$                            &23.37&5.41&1.10 \\
 \  \   6   \  \    & $[(4f^5_{5/2}4f^7_{7/2})_{5/2}5s^2]_4$         & $[((4f^6_{7/2})_65s_{1/2})_{11/2}5p_{3/2}]_7$                           &23.40&7.35&1.60 \\
  \  \          & $[(4f^5_{5/2}4f^7_{7/2})_{5/2}5s^2]_3$         & $[((4f^5_{5/2}4f^7_{7/2})_35s_{1/2})_{5/2}5p_{3/2}]_4$             &23.40&6.14&1.66 \\
 \  \   7  \  \     & $[(4f^6_{7/2})_45s^2]_4$                              & $[((4f^6_{7/2})_45s_{1/2})_{7/2}5p_{3/2}]_5$                             &23.53&5.82&2.69 \\
 \  \   8 \  \      &$[(4f^5_{5/2}4f^7_{7/2})_{5/2}5s^2]_4$          & $[((4f^5_{5/2}4f^7_{7/2})_45s_{1/2})_{7/2}5p_{3/2}]_5$             &23.60&4.09&1.48 \\
 \  \   9  \  \     & $[(4f^6_{7/2})_45s^2]_4$                              & $[((4f^6_{7/2})_45s_{1/2})_{9/2}5p_{3/2}]_3$                             &23.65&6.13&1.80 \\
 \  \   10  \  \   &$[(4f^5_{5/2}4f^7_{7/2})_{5/2}5s^2]_4$          & $[((4f^5_{5/2}4f^7_{7/2})_55s_{1/2})_{11/2}5p_{3/2}]_5$           &23.69&6.20&1.34 \\
        & $[(4f^5_{5/2}4f^7_{7/2})_{5/2}5s^2]_3$         & $[((4f^5_{5/2}4f^7_{7/2})_35s_{1/2})_{7/2}5p_{3/2}]_3$             &23.69&4.34&1.62 \\
        &$[(4f^6_{7/2})_65s^2]_6$                               & $[((4f^6_{5/2})_65s_{1/2})_{13/2}5p_{3/2}]_5$                           &23.70&2.78&1.02 \\
 \  \   11 \  \    & $[(4f^6_{7/2})_65s^2]_6$                              & $[((4f^6_{7/2})_45s_{1/2})_{9/2}5p_{3/2}]_4$                             &23.73&5.73&2.73 \\
        & $[(4f^5_{5/2}4f^7_{7/2})_{5/2}5s^2]_5$         & $[((4f^5_{5/2}4f^7_{7/2})_55s_{1/2})_{9/2}5p_{3/2}]_6$             &23.74&6.84&1.95 \\
 \  \   12 \  \    & $[(4f^4_{5/2})_45s^2]_4$                              & $[((4f^6_{5/2})_45s_{1/2})_{7/2}5p_{3/2}]_5$                             &23.81&7.68&5.72 \\
 \  \   13  \  \   & $[(4f^6_{7/2})_65s^2]_6$                              & $[((4f^6_{5/2})_65s_{1/2})_{11/2}5p_{3/2}]_7$                           &23.88&5.83&3.18 \\
        & $[(4f^5_{5/2}4f^7_{7/2})_{5/2}5s^2]_6$         & $[((4f^5_{5/2}4f^7_{7/2})_65s_{1/2})_{13/2}5p_{3/2}]_5$           &23.90&6.65&2.16 \\
        & $[(4f^5_{5/2}4f^7_{7/2})_{5/2}5s^2]_6$         & $[((4f^5_{5/2}4f^7_{7/2})_65s_{1/2})_{11/2}5p_{3/2}]_7$            &23.92&4.69&1.09 \\
 \  \  14 \  \    &$[(4f^5_{5/2}4f^7_{7/2})_{5/2}5s^2]_4$          & $[((4f^5_{5/2}4f^7_{7/2})_45s_{1/2})_{9/2}5p_{3/2}]_3$             &23.97&6.79&3.08 \\
\hline\hline
\end{tabular}
\end{table*}

\begin{figure*}
\centering\includegraphics{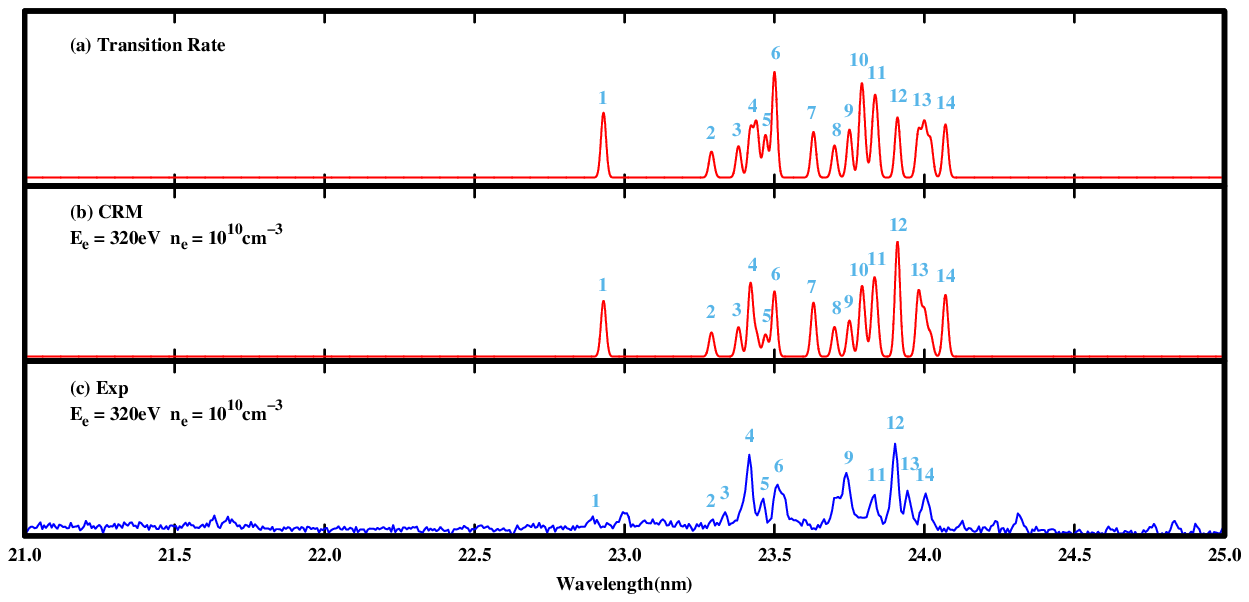}
\caption{Comparison between the experimental and calculated spectra of W$^{14+}$ ion. (a) The calculated radiative transition rate (Shifted to the right by 0.10 nm); (b) The spectral intensity calculated by the CRM (Shifted to the right by 0.10 nm); (c) Experimental spectra measured by Y. Kobayashi et al on the EBIT device with the electron density $n_e = 10^{10}cm^{-3}$ and the energy of electron $E_e = 320eV$.}\label{fig2}
\end{figure*}

The experimental and calculated spectra of W$^{14+}$ ion in the range of 20.0 - 27.0 nm are shown in Fig.~\ref{fig2}. The upper panel Fig.~\ref{fig2} (a) is the calculated radiative transition rate; the middle panel Fig.~\ref{fig2} (b), the spectral intensity calculated by the CRM with the electron density $n_e = 10^{10}cm^{-3}$ and the electron beam energy $E_e = 320eV$; the bottom panel  Fig.~\ref{fig2} (c) is the experimental spectra measured by Y. Kobayashi et al on the EBIT device with the electron density $n_e = 10^{10}cm^{-3}$ and the energy of electron $E_e = 320eV$, and the spectral resolution is 0.03 nm\cite{43}. Each individual transition was assumed to have the Gaussian profile with full width at half maximum (FWHM) of 0.03 nm. The theoretical transition wavelengths are shifted to longer wavelength by 0.10 nm to compare with the experimental spectra. The calculated spectrum of W$^{14+}$ ion is agree with the spectrum of W$^{13+}$ observed by W. Li et al.\cite{42}. The observed intensities of the spectra from W. Li et al. and Y. Kobayashi et al. are different\cite{42,43}. For example, the lines with the key 6 and 11 are strong in W. Li et al.'s work\cite{42}, but weak in Y. Kobayashi et al.'s\cite{43}. The line 13 are weak in W. Li et al.'s work\cite{42}, but strong in Y. Kobayashi et al.'s\cite{43}. The calculated spectrum make general good agreement with both experiments, only a few difference was found in the spectral intensity. For example, the lines with the key 5, 8 and 9 are strong in both experiments, but they are relatively weak in the present calculation. The lines with the key 1, 2, 3, 13 and 14 are strong in the present calculation, but they are weak in the experiment\cite{42,43}. The lines with the key 7 and 10 are strong in the calculated spectrum, but not observed in the experimental spectrum. These difference need to be studied by further work both experimentally and theoretically.

\subsection{The transition and spectrum of W$^{15+}$ ion}

The calculated ground configuration of W$^{15+}$ ion is $4f^{11}5s^2$. The transition wavelength, transition rate and spectra intensity of the W$^{15+}$ ion are shown in Table~\ref{Tab6}. The transition in wavelength at 21.48 - 22.54 nm is dominated by $5p - 5s$ transition from $4f^{11}5s5p$ to $4f^{11}5s^2$. Similar to the spectrum of W$^{14+}$, there are 6 lines with stronger intensity in the experimental spectrum of W$^{15+}$, and there are also more than a dozen of calculated lines with high transition rates. Only the transition data for transitions with high spectra intensity are given in the table. The blended transitions are found in the present calculation. For example, the line with the wavelength 22.70 nm is blended by three components. And the lines with the key 2 and 5 have two components mixed together.

\begin{figure*}
\centering\includegraphics{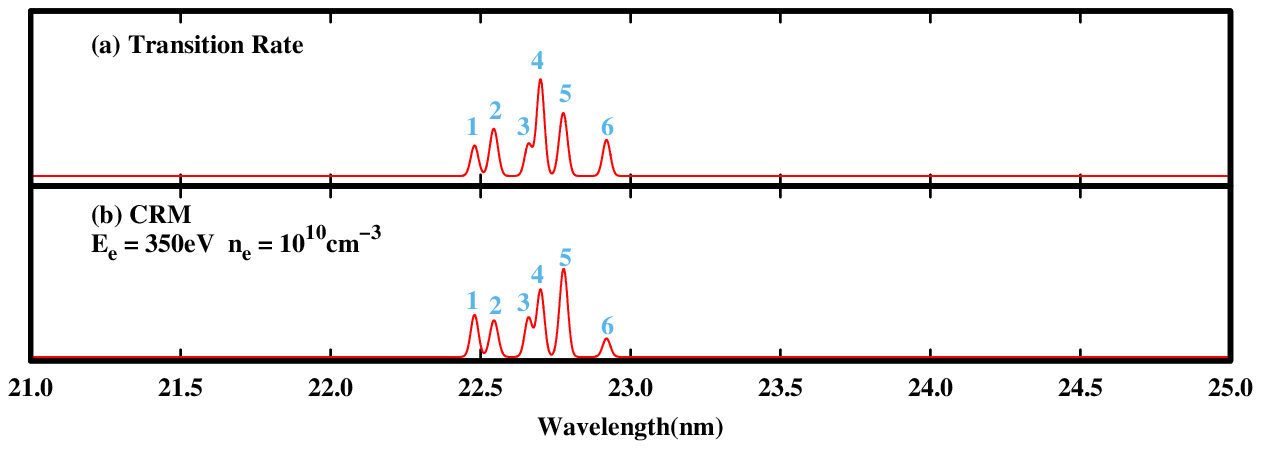}
\caption{The synthetic spectrum of W$^{15+}$ ion. (a) the calculated radiative transition rate; (b) the synthetic spectral calculated by the CRM.}\label{fig3}
\end{figure*}

The synthetic spectrum of W$^{15+}$ ion is shown in Fig.~\ref{fig3}. The upper panel Fig.~\ref{fig3} (a) is the calculated radiative transition rate and the bottom panel Fig.~\ref{fig3} (b) is the spectral intensity calculated by the CRM with the electron density $n_e = 10^{10}cm^{-3}$ and the electron beam energy $E_e = 350eV$. Each individual transition was assumed to have the Gaussian profile with full width at half maximum (FWHM) 0.03 nm. The calculated spectrum of W$^{15+}$ ion is similar to the spectrum of W$^{14+}$ ion observed by W. Li et al. with EBIT\cite{42}. To compare with the experimental spectra of W. Li et al.\cite{42}, the theoretical transition wavelengths are shifted to longer wavelength by 0.20 nm. However, it should be verified by the future experiment.

\begin{table*}
\centering
\caption{The transition wavelength $\lambda$ (in nm), transition rate $A_{(p,q)}$ (in $10^{10}s^{-1}$) and intensity Int  (in $cm^{-3}s^{-1}$) from $4f^{11}5s5p$ to $4f^{11}5s^2$ transitions in $W^{15+}$ ion. The column `Key' correspond to the label in Fig.~\ref{fig3}.}\label{Tab6}
\begin{tabular}{clllll}
\hline\hline
 \  \ Key  \  \ & Lower & Upper & $\lambda$ & $A_{(p,q)}$  & Int \\
\hline
 \  \ 1    \  \    & $[(4f^5_{7/2})_{15/2}5s^2]_{15/2}$                          & $[((4f^5_{7/2})_{15/2}5s_{1/2})_85p_{3/2}]_{15/2}$                   &22.48&5.89&2.36 \\
 \  \ 2    \  \    & $[(4f^5_{7/2})_{11/2}5s^2]_{11/2}$                           & $[((4f^5_{7/2})_{11/2}5s_{1/2})_65p_{3/2}]_{11/2}$                     &22.54&5.59&1.13 \\
        & $[(4f^5_{5/2}4f^6_{7/2})_{13/2}5s^2]_{13/2}$          & $[((4f^5_{5/2}4f^6_{7/2})_{13/2}5s_{1/2})_75p_{3/2}]_{13/2}$    &22.55&4.26&1.09 \\
 \  \ 3    \  \    & $[(4f^5_{7/2})_{15/2}5s^2]_{15/2}$                          & $[((4f^5_{7/2})_{15/2}5s_{1/2})_85p_{3/2}]_{13/2}$                    &22.66&6.20&2.21 \\
 \  \ 4    \  \    & $[(4f^5_{7/2})_{9/2}5s^2]_{9/2}$                              & $[((4f^5_{7/2})_{9/2}5s_{1/2})_45p_{3/2}]_{11/2}$                      &22.70&6.74&1.38 \\
        & $[((4f^4_{5/2})_44f^7_{7/2})_{7/2}5s^2]_{11/2}$    \   \  \   \  \  \  \  \  \  & $[((4f^4_{5/2}4f^7_{7/2})_{11/2}5s_{1/2})_55p_{3/2}]_{13/2}$   \   \  \   \  \  \  \  \  \  &22.70 \  \  \  \  \  \  \  \  &6.83 \  \  \  \  \  \  \  \  &1.14 \  \  \  \\
        & $[(4f^5_{7/2})_{11/2}5s^2]_{11/2}$                          & $[((4f^5_{7/2})_{11/2}5s_{1/2})_55p_{3/2}]_{13/2}$                     &22.70&5.13&1.25 \\
 \  \ 5    \  \    & $[(4f^5_{5/2}4f^6_{7/2})_{13/2}5s^2]_{13/2}$          & $[((4f^5_{5/2}4f^6_{7/2})_{13/2}5s_{1/2})_65p_{3/2}]_{15/2}$     &22.77&5.52&1.67 \\
        & $[(4f^5_{7/2})_{15/2}5s^2]_{15/2}$                          & $[((4f^5_{7/2})_{15/2}5s_{1/2})_75p_{3/2}]_{17/2}$                    &22.78&7.63&3.61 \\
 \  \ 6    \  \    & $[(4f^5_{5/2}4f^6_{7/2})_65s^2]_{15/2}$                 & $[((4f^5_{5/2}4f^6_{7/2})_{15/2}5s_{1/2})_75p_{3/2}]_{17/2}$     &22.92&6.99&1.03 \\
\hline\hline
\end{tabular}
\end{table*}
 
\section{Conclusion}

In this paper, the $5p - 5s$ transition spectra of W XIV - W XVI ions have been calculated by the relativistic configuration interaction (RCI) method and collisional radiative model. The present theoretical results are in good agreement with the experimental results. The identification of the ionization degree from W. Li et al. seems lower by one than the present calculation and  Y. Kobayashi's observation. The spectrum of W XVI has been predicted in the present work and will be observed in the future experiment.

\begin{acknowledgments}
This work was supported by the National Key Research and Development Program (No. 2017YFA0402300) and National Natural Science Foundation of China (No. U1832126, 11874051,11775269) and Users with Excellence Program of Hefei Science Center CAS (No. 2019HSC-UE014).
\end{acknowledgments}

\footnotesize
\bibliography{reference.bib}

\end{document}